\documentclass[prd,onecolumn,nofootinbib,aps,tightenlines,preprintnumbers,notitlepage,longbibliography,superscriptaddress,12pt]{revtex4-1}

\usepackage[dvipsnames]{xcolor}
\usepackage{color}
\usepackage{tikz}
\usepackage[caption=false]{subfig}
\usepackage{multirow}
\usepackage{siunitx}
\usepackage[export]{adjustbox}
\usepackage{graphicx}
\usepackage{dcolumn}
\usepackage{bm}
\usepackage{epstopdf}
\usepackage{tabularx}
\usepackage{suffix}
\usepackage{mathtools}

\usepackage{graphicx}
\usepackage{dcolumn}
\usepackage{bm}
\usepackage{xcolor}

\graphicspath{ {figures/} }

\usepackage{epstopdf}
\newcommand{\be}{\begin{eqnarray}}

\newcommand{\ee}{\end{eqnarray}}

\usepackage[export]{adjustbox}

\def\spinup{\partial\kern-0.3em\raise0.42ex\hbox{\tiny\textbackslash}}
\def\spindown{\overline{\partial\kern-0.3em\raise0.42ex\hbox{\tiny\textbackslash}}}

\usepackage[dvipsnames]{xcolor}
\usepackage{color}
\usepackage{tikz}
\usepackage[caption=false]{subfig}
\usepackage{multirow}
\usepackage{siunitx}
\usepackage[export]{adjustbox}
\usepackage{graphicx}
\usepackage{dcolumn}
\usepackage{bm}
\usepackage{epstopdf}
\usepackage{tabularx}
\usepackage{suffix}
\usepackage{mathtools}
\usepackage{graphicx}
\usepackage{amsmath}
\usepackage{amsfonts}
\usepackage{float}
\usepackage[colorlinks=true,citecolor=blue,urlcolor=cyan,linkcolor=black]{hyperref}

\usepackage{amsmath}
\usepackage{amsfonts}

\usepackage{tabularx}
\newcolumntype{C}{>{\centering\arraybackslash}X}
\newcolumntype{R}{>{\raggedleft\arraybackslash}X}

\usepackage[normalem]{ulem}
\usepackage{xcolor}
\newcommand{\dd}{{\rm d}}

\definecolor{colorA}{HTML}{1E90FF}
\definecolor{colorB}{HTML}{228B22}
\definecolor{colorC}{HTML}{FF7F00}
\definecolor{colorD}{HTML}{4B0082}
\definecolor{colorE}{HTML}{B22222}

\definecolor{lgreen}{HTML}{32CD32}
\definecolor{lgray}{HTML}{D3D3D3}
\definecolor{dblue}{HTML}{1E90FF}
\definecolor{dblue}{HTML}{1E90FF}
\definecolor{orange}{HTML}{FF4500}
\definecolor{indigo}{HTML}{4B0082}

\definecolor{teal}{HTML}{008080}
\definecolor{firebrick}{HTML}{B22222}
\definecolor{salmon}{HTML}{FA8072}
\definecolor{darkgreen}{HTML}{006400}

\newcommand{\jhu}{William H. Miller III Department of Physics and Astronomy, Johns Hopkins University, Baltimore, MD 21218, USA}

\newcommand{\perimeter}{Perimeter Institute for Theoretical Physics, 31 Caroline St N, Waterloo, ON N2L 2Y5, Canada}

\newcommand{\berkeleya}{Lawrence Berkeley National Laboratory, One Cyclotron Road, Berkeley, CA 94720, USA}
\newcommand{\berkeleyb}{Berkeley Center for Cosmological Physics, Department of Physics, University of California, Berkeley, CA 94720, USA}

\newcommand{\usc}{Physics \& Astronomy Department, University of Southern California, Los Angeles, California, 90089-0484}

\graphicspath{ {plots/} }

\begin{document}

\title{Transverse velocities and matter gradient correlations: a new signal and a new challenge to moving-lens analyses}

\author{Selim~C.~Hotinli}
\email{shotinl1@jhu.edu}
\affiliation{\jhu}

\author{Elena~Pierpaoli}
\email{pierpaol@usc.edu}
\affiliation{\usc}

\author{Simone~Ferraro}
\email{sferraro@lbl.gov}
\affiliation{\berkeleya}
\affiliation{\berkeleyb}

\author{Kendrick~Smith}
\email{kmsmith@perimeterinstitute.ca}
\affiliation{\perimeter}

\begin{abstract}
{An observer that is moving towards a high-density region sees,  on average, a higher matter density and more foreground-emitting sources ahead than behind themself. Consequently, the average abundance and luminosity  of objects producing cosmological signals around an in-falling dark matter halo  is larger in the direction of the halo’s motion. In this Letter, we demonstrate this effect from simulated cosmological maps of the thermal Sunyaev Zel’dovich effect and the cosmic infrared background. We find that, for a wide range of halo masses and redshifts, oriented stacked profiles of these foregrounds show significant, potentially detectable gradients aligned with the transverse velocity of halos. The signal depends on the halo’s mass and redshift, as well as the physical properties of the cosmic web surrounding the halos. We show that this signal is sufficiently prominent to be detected in future Cosmic Microwave Background experiments, therefore offering a new window into the study of cosmological structures. We argue that the dipolar morphological structure of this signal, its orientation,  as well as  its overall large amplitude, constitute a challenge for the detection of the transverse velocity through the study of the  moving lens effect for stacked halos.}
\end{abstract}
\maketitle
\section{Introduction}

Large-scale structure (LSS) evolves hierarchically, in time forming a network of filaments, super-clusters and virialised objects such as dark matter halos and galaxies~\citep[e.g.][]{Davis:1985rj,Bond:1990iw,Bernardeau:2001qr,Springel:2005nw,Springel:2017tpz}. Most massive halos form at high-density regions that are connected by filaments of dark matter and see an accretion of mass in gas and smaller-mass halos~\citep[e.g.][]{Bond:1995yt}. For observers moving towards a high-density region (in-falling), structure in the local environment would in average appear more clustered in the direction of the peculiar velocity, towards the higher-density region. Equivalently, a cosmologist observing the LSS would expect to see more structure (and hence an enhancement of cosmological signals) near and in front of halos that are falling into higher-density regions. The same effect can also be understood from linear theory, where the velocity field is proportional to gravitational potential gradient. The gravitational potential is strongly correlated with foregrounds (as the foreground emission comes from matter), so the velocity is also in the direction of foreground density gradients. Here we demonstrate this effect from simulations of thermal Sunyaev Zel'dovich (tSZ) effect in the cosmic microwave background (CMB)~\citep{Zeldovich:1969ff,1972CoASP...4..173S,Sunyaev:1980vz} and the cosmic infrared background (CIB)~\citep{hauser2001cosmic,chary2001interpreting,lagache2005dusty}, and propose this enhancement along the transverse velocities ahead of in-falling cosmological structure as a potential observable of local matter density of high density regions. 

Joint analyses of CMB and LSS open new windows of opportunity for cosmological inference~\citep[e.g.][]{Deutsch:2017ybc,Smith:2018bpn,Munchmeyer:2018eey,Zhang:2015uta,Hotinli:2019wdp,Cayuso:2019hen,Alvarez:2020gvl, Hotinli:2020csk, Hotinli:2019wdp,Cayuso:2019hen, Alvarez:2020gvl, Hotinli:2020csk, Foreman:2022ves, Hotinli:2022jna, Lee:2022udm, Hotinli:2022jnt,Hotinli:2021hih,Hotinli:2018yyc,Hotinli:2019wdp,Hotinli:2020ntd,AnilKumar:2022flx,Kumar:2022bly,Hotinli:2022jna,Hotinli:2022jnt}. These programs are increasingly gaining attention as they often increase the prospects to detect and characterise new signals by reducing systematics, cancelling cosmic variance and breaking degeneracies. The CMB signal contains valuable information about LSS, as the CMB photons interact with LSS on their trajectories and source a variety of signals. One such signal is the tSZ effect due to inverse-Compton scattering of CMB photons off on energetic electrons in the circumgalactic medium. Another such signal is the CIB, i.e.\ thermal radiation of dust grains in distant star-forming galaxies. Dust grains absorb the ultraviolet starlight, heat up and re-emit light in the infrared. As star formation rate of our Universe peaks at around $z\sim2-3$, CIB is sourced dominantly from galaxies around these redshifts.

The tSZ and CIB signals correspond to the dominant contribution to the extra-galactic foreground contamination in the CMB, obscuring our view of the pristine primordial CMB fluctuations from the early Universe on small scales. Nevertheless these signals are also a valuable source of cosmological and astrophysical information and constitute to significant science drivers for the upcoming CMB experiments such as Simons Observatory~\citep{2018arXiv180807445T,Abitbol:2019nhf} and CMB-S4~\citep{CMB-S4:2016ple,Abazajian:2019eic}. {Here, we show these signals as measurable probes of the cosmological structure along the direction of halo transverse velocities. This paper is organised as follows: In Section~\ref{sec:foreground_grad}, we show the foreground gradients along in-falling dark matter halos using an oriented stacking method. In Section~\ref{sec:moving_lens} we perform a theoretical study on the relation between the foreground gradients from single-frequency maps and the moving-lens effect. We conclude in Section~\ref{sec:DISCUSSION} with discussion.}

\section{Foreground gradients}\label{sec:foreground_grad}

In order to demonstrate the anisotropic enhancement of structure around clustering halos, we construct an oriented stacking algorithm that selects CMB patches around halos, rotates them to align the patches along the transverse velocity on the 2-sphere, and maps them onto a $N\times N$ grid of pixels, whose centers evenly cover the range $[-\lambda r/r_s,,+\lambda r/r_s]$ in two orthogonal directions on the 2-sphere. Here $r$ is the physical distance from the halo center, {$r_s$ is the scale radius of a given halo defined with $r_s=r_{\rm vir}/c$ where $r_{\rm vir}$ is the radius within which the mean density is 200 times the background density and $c$ is the concentration parameter, for which we assume the standard NFW dark-matter halo model~\citep{Navarro:1995iw},} and $\lambda$ determines the size of the stacked patches in the sky\footnote{For the standard NFW model and LCDM cosmology assuming latest Planck measurements, the scale radius of a $10^{13}$ solar mass halo at redshifts $z=\{0.5,1.,2.\}$ correspond approximately to $\{0.17',0.14',0.13'\}$. For $10^{12}$ and $10^{14}$ solar mass halos, $r_s$ is roughly $\{0.07',0.05',0.05'\}$ and $\{0.44',0.27',0.24'\}$, respectively. }. Once rotated, we stack these $N\times N$ grid of pixels over a catalog of halos. We detail our stacking algorithm in our upcoming paper~\citep{hotinli_elena_2023}.

We apply our stacking algorithm to \texttt{websky} dark matter halo catalog\footnote{\href{https://mocks.cita.utoronto.ca/data/websky/v0.0/}{mocks.cita.utoronto.ca/data/websky/v0.0/}} and extragalactic CMB simulations~\citep{Stein:2020its}. The  \texttt{websky} catalog is modelled with ellipsoidal collapse dynamics and the corresponding displacement field is modelled with Lagrangian perturbation theory. The catalog spans a redshift interval $0<z<4.6$ over the full sky, spanning a volume of $\sim600~({\rm Gpc}/h)^3$ and consists of approximately a billion halos of  mass satisfying $M_h>10^{12} M_\odot$ where $M_\odot$ is the solar mass. The \texttt{websky} halo catalog and the displacement field is also used to generate a range of intensity maps to simulate scattering and lensing effects on the CMB photons. These publicly available maps include the tSZ and CIB effects, as well as the kinetic Sunyaev Zel'dovich effect (kSZ) {induced by scattering from free electrons in bulk motion and} the weak gravitational lensing of the CMB due to intervening large-scale structure. Here, we demonstrate the enhancement of structure ahead of infalling halos using the tSZ and CIB maps from \texttt{websky}. Throughout this letter, we show this effect with maps of tSZ and CIB at frequencies $150\,{\rm GHz}$ and $220\,{\rm GHz}$, respectively. The maps we use for demonstration in Figs.~(1-3) have $\sim0.8'$ resolution (\texttt{healpix}\footnote{\href{https://healpix.sourceforge.io/}{https://healpix.sourceforge.io/}} format, NSIDE=4096). 

\begin{figure*}[t!]
\centering
\includegraphics[width=1.025\linewidth]{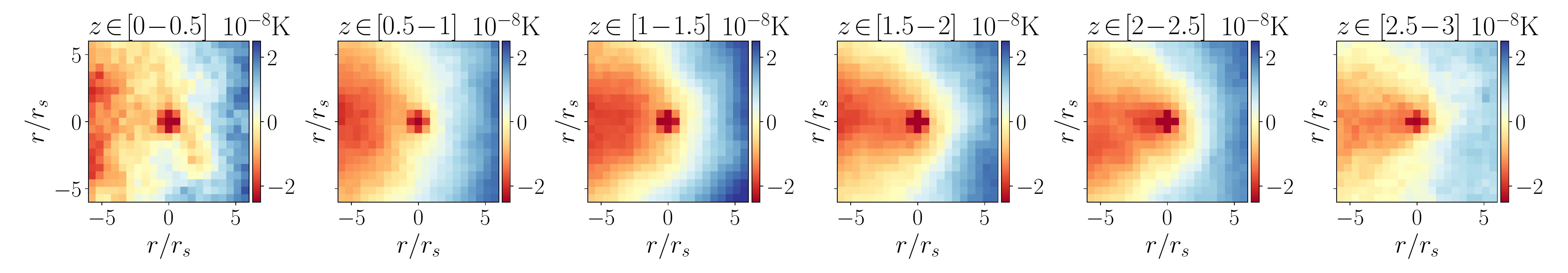}\vspace*{-0.2cm}
\includegraphics[width=1.025\linewidth]{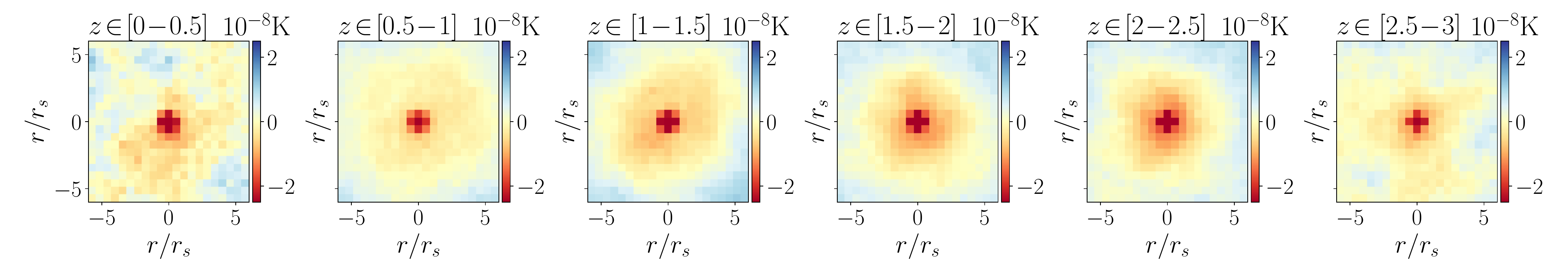}\vspace*{-0.2cm}
\includegraphics[width=1.025\linewidth]{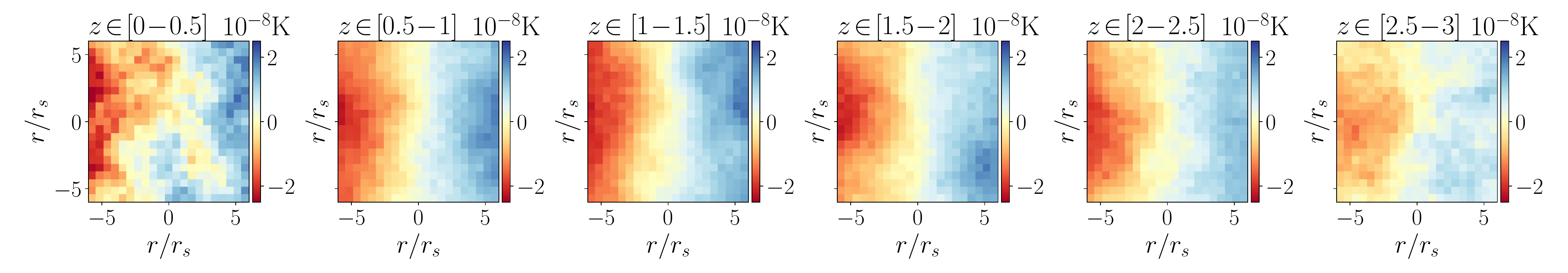}\vspace*{-0.5cm}
\caption{Oriented stacks of tSZ signal at frequency $\nu=150\,{\rm GHz}$. Each patch is taken around halos in the \texttt{websky} simulation satisfying the mass range $M_{\rm h}\simeq\,10^{12}M_{\odot}$ and within the ranges of redshifts shown above each panel. The top row of panels correspond to stacks of the tSZ signal from patches oriented along the true velocities of halos in the simulation (along the $x$-axis, to the left).  The gradient of the signal along the $x$-axis can be seen clearly clearly in most of the stacks. The middle row of panels corresponds to stacks of patches oriented randomly, where the gradient is absent. The lower row of panels shows the difference between the two panels, i.e.\ between stacked patches oriented along the true transverse velocities  of corresponding halos and those that are oriented randomly. Values of pixels at each panel correspond to the mean temperature of stacked patches in Kelvin. The density gradient along the direction of transverse velocity shows an increased average tSZ signal ahead of the halos, apparent at all redshifts. The gradient becomes less apparent both at lowest $z\in[0-0.5]$ and highest redshifts $z\in[2.5-3.0]$ we consider, as we discuss in the text.}
\label{fig:FIG1}
\end{figure*}

\begin{figure*}[t!]
\centering
\includegraphics[width=1.025\linewidth]{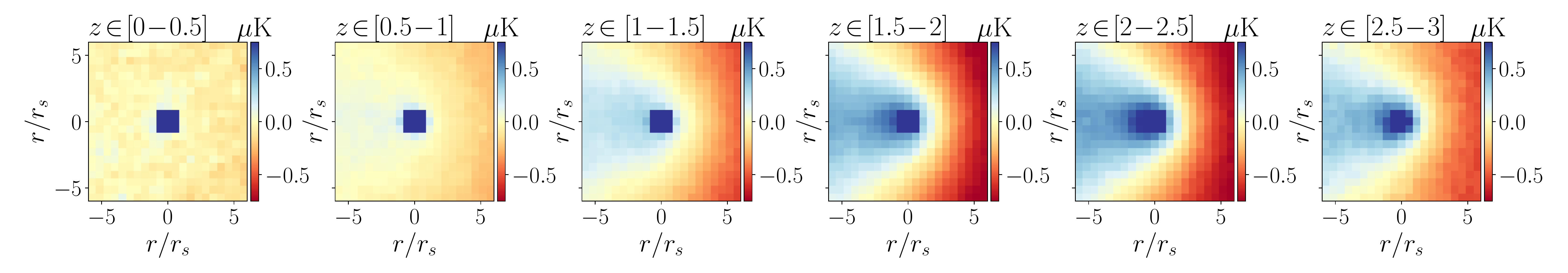}\vspace*{-0.2cm}
\includegraphics[width=1.025\linewidth]{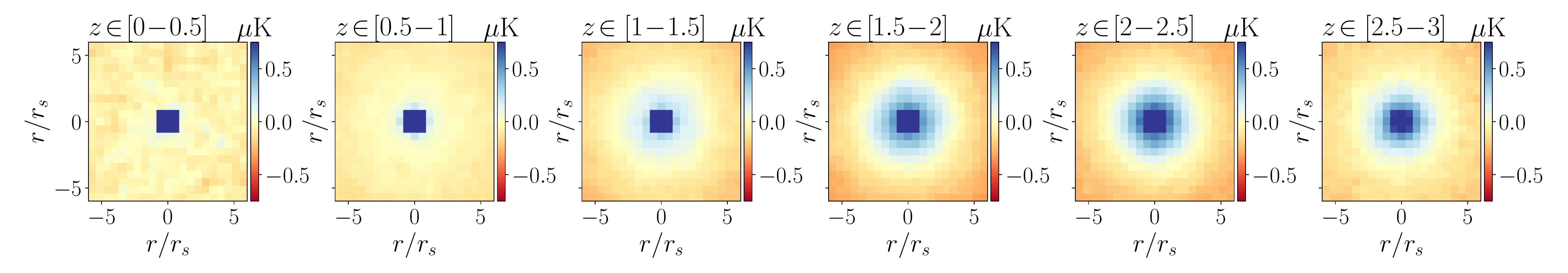}\vspace*{-0.2cm}
\includegraphics[width=1.025\linewidth]{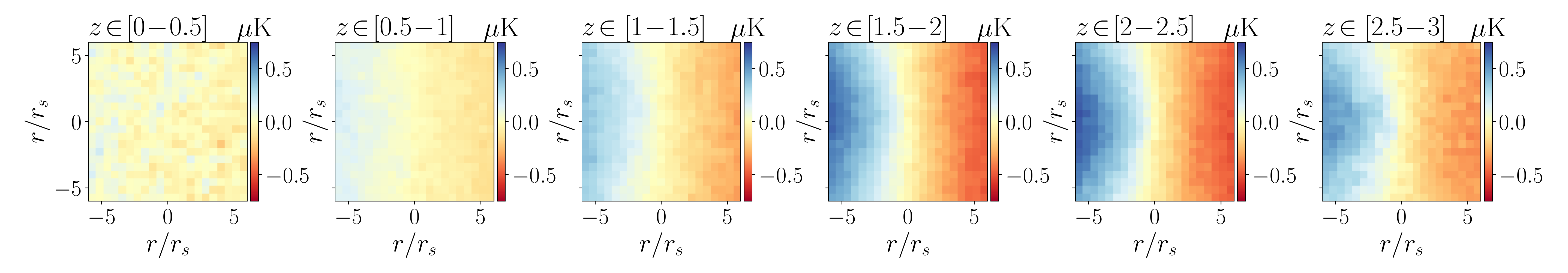}\vspace*{-0.5cm}
\vspace*{-0.5cm}
\caption{Oriented stacks of CIB signal at frequency $\nu=220\,{\rm GHz}$. Figure otherwise identical to to the lowest row of panels in Fig.~\ref{fig:FIG1}. Note that as the CIB signal is sourced at redshifts $z\in[1-4]$, the gradient in the patches become observable for redshifts satisfying $z\gtrsim1$.}
\label{fig:FIG2}
\end{figure*}\vspace*{-0.1cm}

Figures~\ref{fig:FIG1}~and~\ref{fig:FIG2} show results from stacking $\sim10$ million tSZ and CIB patches, distributed over the full sky, around halos of mass $M_{h}\sim\,10^{12}M_\odot$, at frequencies $\nu=150\,{\rm GHz}$ and $\nu=220\,{\rm GHz}$, respectively. The top row of panels in Fig.~\ref{fig:FIG1} corresponds to stacks from patches oriented along the transverse velocity of the halos (along the $x$-axis, to the left). The middle row of panels in the same figure corresponds to stacks from patches oriented randomly. The foreground gradient can be seen from the top row of panels, which is absent in the middle row. The bottom row in the same figure corresponds to the \textit{difference} between the top and the middle panels. For all stacks, the redshift range of halos of each patch is labelled at the top of the panels. The pixel values (in $\mu$K) correspond to the mean temperature of the patches in that stack. The enhanced signal along the direction of the transverse velocity is apparent on all stacks shown in the bottom row, and is most pronounced at redshift bins spanning the range $z\in[0.5-2.0]$, where the contribution to the tSZ signal from halos of mass $M_{h}\sim\,10^{12}M_\odot$ is most significant. Fig.~\ref{fig:FIG2} corresponds to the same analysis performed on maps of CIB, where redshift and mass choices are otherwise identical. As the the CIB signal is sourced at higher redshifts, the gradient becomes apparent only for stacks composed of halos at redshifts $z\gtrsim1$.

The mass dependence of the foreground gradients are shown in Fig.~\ref{fig:FIG3}. The left (right) three columns correspond to tSZ (CIB) signals, similarly at frequency 150GHz (220GHz). The top row of columns corresponds to stacks from halos within the redshift range $z\in[0,0.5]$ while middle and bottom rows of panels correspond to redshift ranges $z\in[1.4,1.8]$ and $z\in[2.5,3.0]$, respectively. The gradients can be clearly seen on the tSZ stacks at all redshifts for halos of masses $M_h\simeq10^{12}M_\odot$ and $\simeq10^{13}M_\odot$. At the highest range of redshifts we consider, oriented stacking around the most massive halos $M_h\simeq10^{14}M_\odot$ do not lead to a foreground gradient, as at high redshifts these halos constitute to the central, most massive halos in the overdensities (hence they are not in-falling). At low redshifts, halos up to $M_h\sim10^{14}\,M_\odot$ appear to be still falling into denser environments as evident in the tSZ stacks. The shape of the enhancement of the foregrounds along the direction of velocity can be seen to be dependent on the halo mass and redshift. For lower mass halos, the peak of the enhancement falls further from the center of the stacks, in a way dependent on the difference between the scale radius of the stacked halo and the central, more massive halos in that overdensity. The same results apply for the stacks of CIB with the noteable difference that the CIB signal {is not} sourced on low redshifts, hence the foreground gradient becomes apparent only for the middle and bottom panels. The number of halos in the stacked patches shown in this Figure are approximately $10^{6},10^{5},10^{4}$ for masses $\simeq10^{12}$, $\simeq10^{13}$ and $\simeq10^{14}$, respectively, for both tSZ and CIB.

\section{Relation to the moving-lens signal}\label{sec:moving_lens}

\begin{figure*}[t!]
\centering
\includegraphics[width=1.025\linewidth]{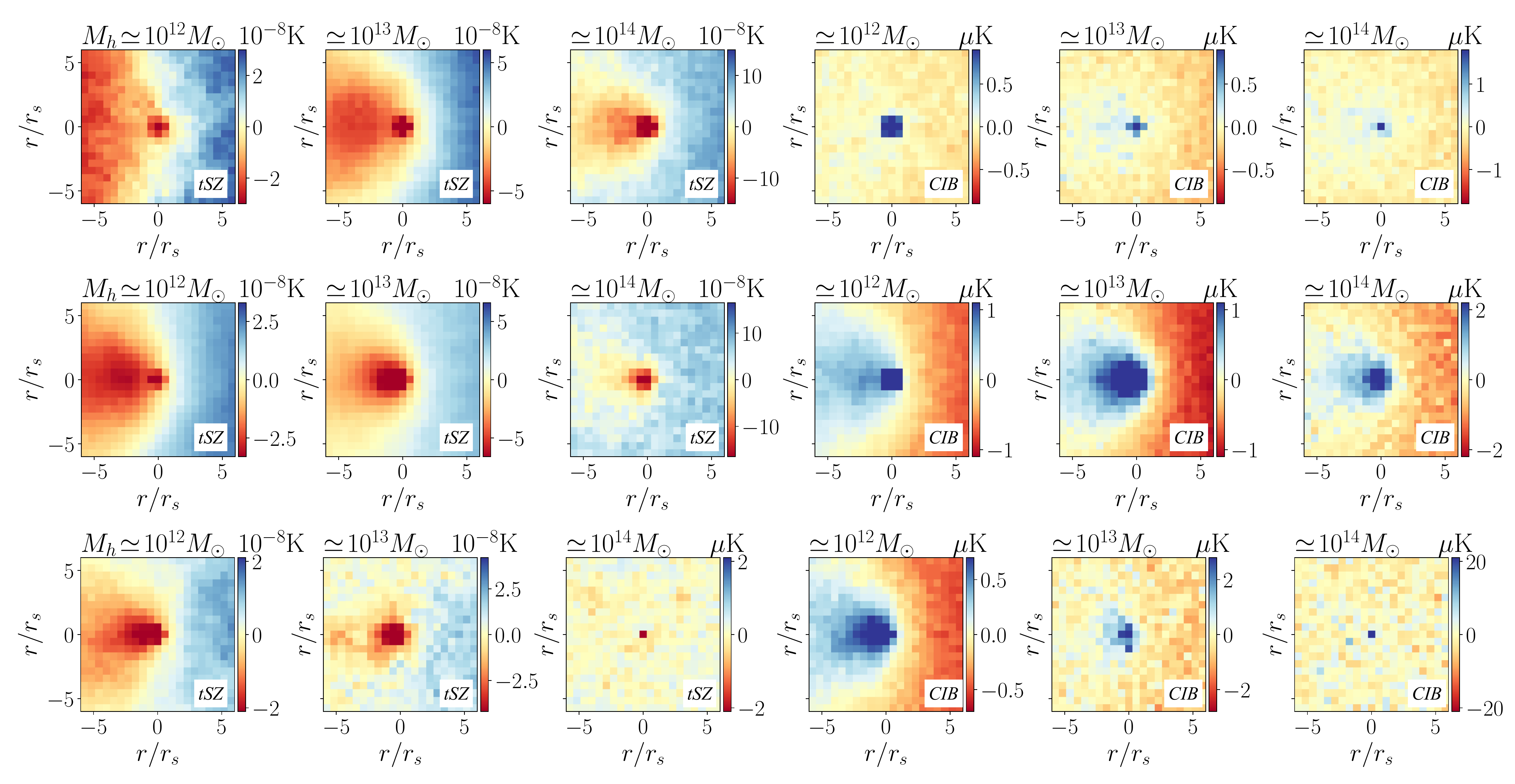}\vspace*{-0.2cm}
\caption{{Oriented stacks of tSZ maps at $150$GHz (left panels) and CIB maps at $220$GHz (right panels)}, around halos of varying mass. The top row of panels corresponds to halos within the redshift range $z\in[0.0,0.5]$. The middle row of panels corresponds to halos satisfying $z\in[1.4,1.8]$. Lowest row of panels corresponds to halos satisfying $z\in[2.5,3.0]$. The approximate mass of the halos for each stack is shown above the panels. For the highest redshift range (lowest row of panels), the gradients are apparent only for the lowest mass halos. As the CIB signal is sourced most dominantly at redshifts $z\sim2$, the middle row of panels show a more apparent gradient for most halo masses. We discuss the physics leading to these expected characteristics in the text.}\vspace*{-0.2cm}
\label{fig:FIG3}\vspace*{-0.3cm}
\end{figure*}

The redshift and mass dependence of the CIB and tSZ foreground gradients suggest observations of this signal can be useful for probing both the linear and non-linear structure formation, which we return later in Section~\ref{sec:DISCUSSION}. {This signal can also be a potential nuisance} for measuring the moving-lens effect, a specific dipolar signature aligned with the transverse velocity of halos~\citep{Hotinli:2018yyc, Yasini:2018rrl}. The moving-lens effect is a Doppler effect of the CMB photons that get redshifted or blueshifted upon traversing gravitational potentials that are moving transverse to the line of sight. Upcoming CMB experiments are expected to make this detection in the near future for the first time in cross-correlation with galaxy surveys.

\begin{figure}[H]
\centering
\includegraphics[width=0.6\linewidth]{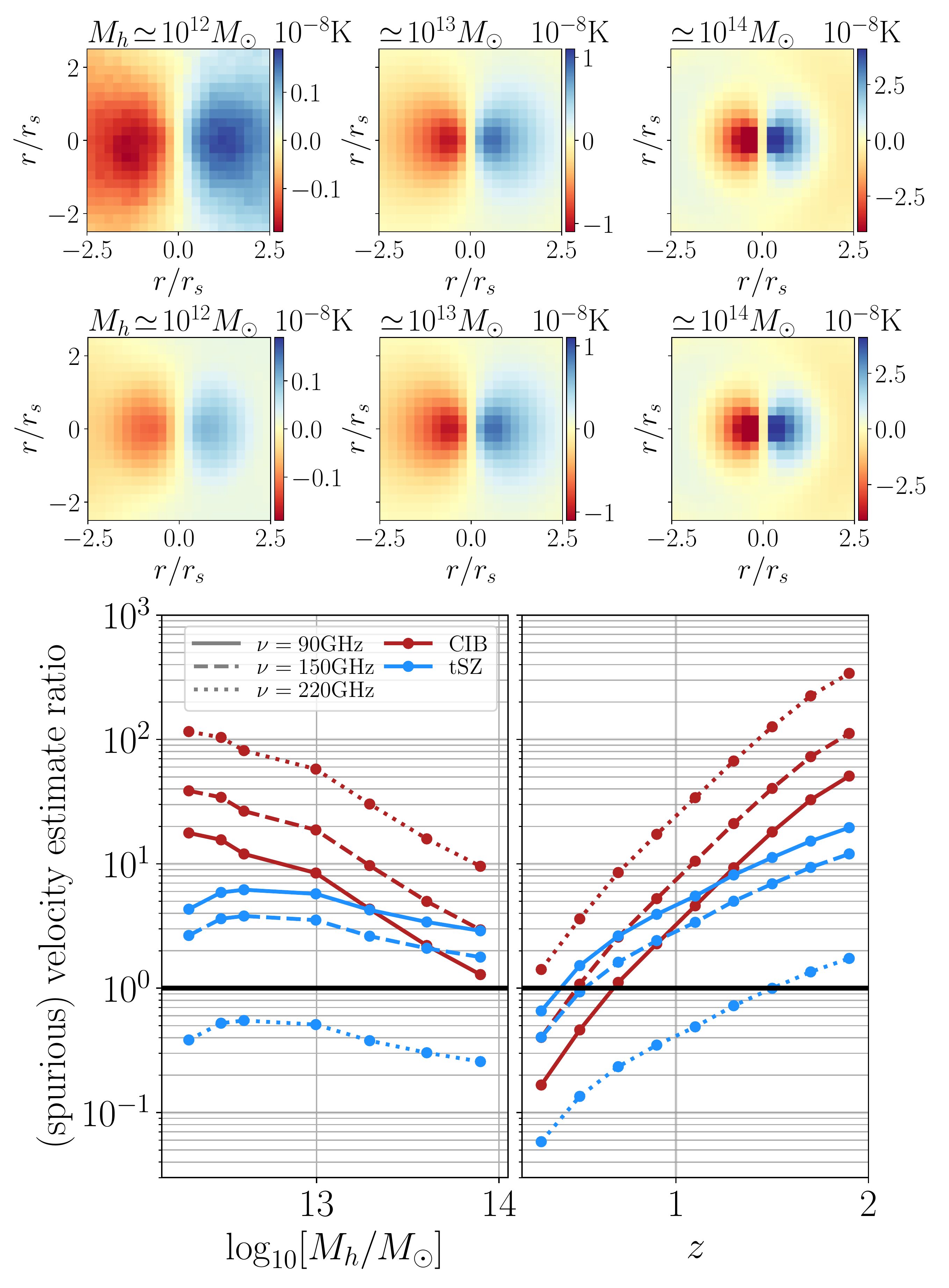}\vspace*{-0.3cm}
\caption{\textit{{The comparison between the moving-lens effect and \textit{single-frequency} CIB and tSZ maps.}} The upper ($2\times3$) panels correspond to stacked profiles from a map of the moving lens effect generated using the \texttt{websky} simulation. The first and second rows correspond to stacks of halos within the redshift ranges $z\in[0,0.5]$ and $z\in[1.4,1.8]$, respectively. The stacks of the moving lens effect show a dipolar pattern that is expected to be detected in with upcoming CMB surveys in the near future. {In order to acquire a theoretical understanding of the potential effect of foreground gradients to velocity estimation, here we use a matched filter appropriate for the anticipated $1\mu$K' noise and $1.4'$ beam of the upcoming CMB-S4 survey at single frequencies to estimate the transverse peculiar velocity from the moving-lens-only signal. We describe our filter and the CMB spectrum we use in more detail in Sec.~\ref{sec:moving_lens}. We apply the same filter to single-frequency CIB and tSZ maps. The lower two panels show the ratio of the spurious velocity estimates from single-frequency CIB and tSZ maps to velocity estimates from the moving-lens effect.  We find single-frequency CIB and tSZ foreground maps lead to velocity estimates that are around an order magnitude larger than the signal, which may indicate a significant challenge for moving-lens effect detection. The left and right lower panels demonstrate the mass and redshift dependence of this effect. We find the spurious velocities from lowest-mass higher-redshift (higher-mass lower-redshift) stacks to be most (least) significant in comparison. In practice, stacks will include other signals including the lensed (or delensed, see e.g.~Ref.~\citep{Hotinli:2021umk}) primary CMB and survey-specific and frequency-dependent noise, for example, and ILC-cleaning will reduce the net contamination from CIB and tSZ foregrounds.}}\vspace*{-0.55cm}
\label{fig:FIG4}
\end{figure}

{The upper ($2\times3$) panels in Fig.~\ref{fig:FIG4} show stacks from maps consisting of moving-lens effect only, produced from the same \texttt{websky} catalogue. The dipolar structure of this signal has similarities with the tSZ and CIB showed in Figs~(\ref{fig:FIG1}-\ref{fig:FIG3}), but it is overall much smaller.  
It is therefore expected that the tSZ and CIB signals will impact the detection and measurement of the moving-lens effect.
Ultimately,  component separation methods, like the standard `Internal Linear Combination' (ILC),  can reduce the contamination from frequency-dependent foregrounds (such as CIB and tSZ) by combining information from various  frequency channels.} However, some foreground residuals are expected in the component-separated map. Here we take a simple approach to gain a theoretical understanding of how impactful the CIB and tSZ  effects could be, if untreated, on the moving-lens stacked images (on the determination of peculiar velocities from the moving-lens stacked images). 

We apply a filter to stacked single frequency maps which is optimized for detecting the moving lens effect and to determine the transverse velocity sourcing the effect~\citep{Hotinli:2020ntd}.
The simplest version of the filter is designed to be optimal and produce unbiased results in case no foregrounds are present, and the maps only contain the moving lens signal, the CMB and instrumental noise.  
We apply a generalized version of this filter to  simulated single-frequency maps of CIB and tSZ, as well as  a map of the moving-lens effect. Here we apply a filter designed for a CMB-S4 like survey with the Fourier-space shape $$\tilde{\Psi}(\vec{\ell})={N}^{\rm rec}\frac{\tilde{\mathcal{M}}(\vec{\ell}|M,z)}{\tilde{C}^{TT}_\ell}$$ where $\tilde{\mathcal{M}}(\vec{\ell}|M,z)$ is the Fourier-space profile of the moving-lens signal given halo mass and redshift, aligned with the transverse velocity of the halo, and $\tilde{C}^{TT}_\ell$ is the lensed CMB temperature power-spectrum (including single-frequency tSZ, CIB foregrounds, the late-time and reionization kSZ effect as well as radio sources as described in~\citep{Hotinli:2021hih}) with CMB-S4 specific $1.4'$ beam and $\sim1\mu{\rm K}'$ residual white noise appropriate for a single cutout. The moving-lens, CIB and tSZ maps we use are filtered with the same $1.4'$ beam. The estimator variance satisfies $$N^{\rm rec}=a_0^{-2}\!\!\int \frac{\dd^2\vec{\ell}}{(2\pi)^2} {|\tilde{\Psi}(\vec{\ell})|^2}{\tilde{C}_\ell^{TT}}$$ where $a_0={16\pi G\rho_sr_s^2}{c^{-3}}$.

Our results are shown in Fig.~\ref{fig:FIG4} (lower panels). 
We find the amplitude of the spurious velocity estimates from the CIB and tSZ maps to be significant (an order of magnitude larger) compared to the velocity estimates from the moving-lens effect. 
The observed bias depends on masses and redshifts of the stacked halos. Our results clearly show that upcoming cosmological applications of the moving lens effect will be largely dependent on how well the gradients of CIB and tSZ will be cleaned by the component separation technique adopted. 

A tailored component-separation technique may leverage frequency dependencies of the various signals (as e.g.\ in the standard ILC) but also on morphological differences of the moving-lens from the foregrounds. 
The foreground gradients from single-frequency CIB and tSZ maps  have a different spatial morphology than the moving lens signal.
The largest contribution to the tSZ and CIB foregrounds away from the halo centers, for example, can be seen to peak at farther distances from the halo center for lower mass halos. In addition, in particular for the CIB foreground, the gradients are absent for stacks of low-redshift halos, as the signal comes from high redshifts. These unique characteristics can potentially be utilized to overcome the bias on the moving-lens analyses, once realistic survey-specific considerations are taken into account. Nevertheless, our  results warrant further study and forecasts on the detection prospects of the moving-lens effect taking into account this potential bias from residual CIB and tSZ on ILC-cleaned maps of CMB, which was previously unnoticed. We perform these analyses in our upcoming paper on this subject~\citep{hotinli_elena_2023}.

\section{Discussion and Conclusions}\label{sec:DISCUSSION}

Foreground and matter gradients aligned with transverse velocities--and hence the enhancement of cosmological signals including CMB foregrounds ahead of in-falling halos--is an expected outcome of standard structure formation. Here, we demonstrated this effect from simulations of tSZ and CIB signals. Our analysis is limited to making the observation that the oriented stacked patches of these foregrounds see a clearly-noticeable gradient aligned along the transverse velocities of halos. This effect can be thought as both a new signal, as well as a previously unknown bias to the analysis of the moving-lens effect.

Measurements of cosmological velocities have significant value for cosmological inference as they can be used to probe dark energy, modified gravity and the effects of neutrino mass, for example. In upcoming works, we assess the detectability of the CIB and tSZ foreground gradients with existing and upcoming data and demonstrate the potential scientific value of such measurements~\citep{hotinli_etal}, showing, for example, whether the deviation of these foreground gradients from the linear-theory prediction can provide an important insight into the growth of large-scale structure throughout cosmic time.

In~Ref.~\citep{hotinli_elena_2023}, we also assess the potential bias from this effect on the measurement of the moving lens signal with survey-specific forecasts, taking into account realistic CMB noise, non-Gaussian correlated foregrounds and ILC-cleaning. Different mass and redshift dependence of the CIB and tSZ gradients compared to the moving-lens signal, their different profiles as well as the fact that the tSZ and CIB gradients are in opposite directions at CMB frequencies hence they partly cancel, can potentially be utilized to circumvent these biases--the analysis of which we leave for upcoming work.

\acknowledgements

We thank Marc Kamionkowski, Neal Dalal, Matthew Johnson and Niayesh Afshordi for useful discussions. SCH is supported by the Horizon Fellowship at Johns Hopkins University. KMS was supported by an NSERC Discovery Grant. This research was supported in part by Perimeter Institute for Theoretical Physics. Research at Perimeter Institute is supported by the Government of Canada through the Department of Innovation, Science and Economic Development Canada and by the Province of Ontario through the Ministry of Research, Innovation and Science.

\bibliography{bibfile}
\end{document}